\pdfoutput=1
\documentclass{elsart}

\usepackage{natbib}

\usepackage{graphicx}
\usepackage{color}

\usepackage{amssymb}

\begin{document}

\begin{frontmatter}



\title{Effect of pressure on the superconducting critical temperature of La[O$_{0.89}$F$_{0.11}$]FeAs and Ce[O$_{0.88}$F$_{0.12}$]FeAs}


\author[ucsd]{D.\ A.\ Zocco}
\author[ucsd]{J.\ J.\ Hamlin}
\author[ucsd]{R.\ E.\ Baumbach}
\author[ucsd]{M.\ B.\ Maple\corauthref{mbm}}
\author[or]{M.\ A.\ McGuire}
\author[or]{A.\ S.\ Sefat}
\author[or]{B.\ C.\ Sales}
\author[or]{R.\ Jin}
\author[or]{D.\ Mandrus}
\author[llnl]{J.\ R.\ Jeffries}
\author[llnl]{S.\ T.\ Weir}
\author[bama]{Y.\ K.\ Vohra}
\corauth[mbm]{Corresponding author.\\
		Email address: mbmaple@ucsd.edu}

\address[ucsd]{Department of Physics and Institute for Pure and
Applied Physical Sciences,\\ University of California at San Diego,
La Jolla, CA 92093, USA}
\address[or]{Materials Science \& Technology Division, Oak Ridge
National Laboratory,\\ Oak Ridge, Tennessee 37831, USA}
\address[llnl]{Lawrence Livermore National Laboratory, Livermore,
California 94551, USA}
\address[bama]{Department of Physics, University of Alabama at
Birmingham,\\ Birmingham, Alabama 35294, USA}

\begin{abstract}
We have performed several high-pressure resistivity experiments on the recently discovered superconductors La[O$_{0.89}$F$_{0.11}$]FeAs and Ce[O$_{0.88}$F$_{0.12}$]FeAs. At ambient pressure, these materials have superconducting onset temperatures $T_c$ of 28 K and 44 K, respectively. While the $T_c$ of La[O$_{0.89}$F$_{0.11}$]FeAs goes through a maximum between 10-68 kbar, in qualitative agreement with a recent report by Takahashi \textit{et al.}, the $T_c$ of Ce[O$_{0.88}$F$_{0.12}$]FeAs decreases monotonically over the measured pressure range. At 265 kbar, the $T_c$ of the cerium-based compound has been suppressed below 1.1 K.
\end{abstract}

\begin{keyword}
high pressure \sep designer diamonds \sep superconductivity \sep
oxypnictide
\PACS 74.62.Fj

\end{keyword}

\end{frontmatter}


\section{Introduction}
A new class of superconductors consisting of layered materials with the chemical formula LnOTPn, where Ln is a lanthanide element, T is a transition metal, and Pn is either P, As, or Bi, has recently emerged. The phosphorus-based versions of these compounds, LaOFeP and LaONiP have rather low superconducting critical temperatures, $T_c$ of 3 \cite{watanabe_2007_1} and 5 K \cite{kamihara_2006_1}, respectively. Much higher $T_c$ values were achieved by fluorine-doping the corresponding arsenic-based compound to produce La[O$_{1-x}$F$_x$]FeAs, where doping to $x\sim$ 0.11 produces $T_c$ $\sim$ 26 K \cite{kamihara_2008_1}. The $T_c$ appears to pass through a maximum as a function of fluorine doping. Subsequently, it was found that under a modest pressure of 40 kbar, the $T_c$ of La[O$_{0.89}$F$_{0.11}$]FeAs increases to 43 K \cite{takahashi_2008_1}, becoming the first non-cuprate superconductor with a $T_c$ higher than that of MgB$_2$.  Replacing lanthanum with heavier rare-earth elements also leads to high $T_c$ values, as in Ce[O$_{1-x}$F$_x$]FeAs with $T_c$ up to 41 K \cite{chen_2008_1}.  As of this writing, the highest $T_c$ reported for this class of materials is about 55 K, which was achieved in the compound Sm[O$_{1-x}$F$_x$]FeAs \cite{ren_2008_1}. The $T_c$ of optimally doped Sm[O$_{1-x}$F$_x$]FeAs initially decreases with pressure \cite{lorenz_2008_1}.

In this letter, we report measurements of $T_c$ as a function of pressure in the lanthanum- and cerium-based oxypnictide compounds to further probe the superconductivity in these interesting materials. In Ce[O$_{0.88}$F$_{0.12}$]FeAs, we found that $T_c$ decreases monotonically from 47 K (resistivity onset) to 4.5 K at 190 kbar. In La[O$_{0.89}$F$_{0.11}$]FeAs, $T_c$ appears to pass through a maximum between 10 and 68 kbar, in qualitative agreement with the results of Takahashi \textit{et al.} \cite{takahashi_2008_1}.

\begin{figure}[h!tbp]
    \begin{center}
        \includegraphics[width=8cm]{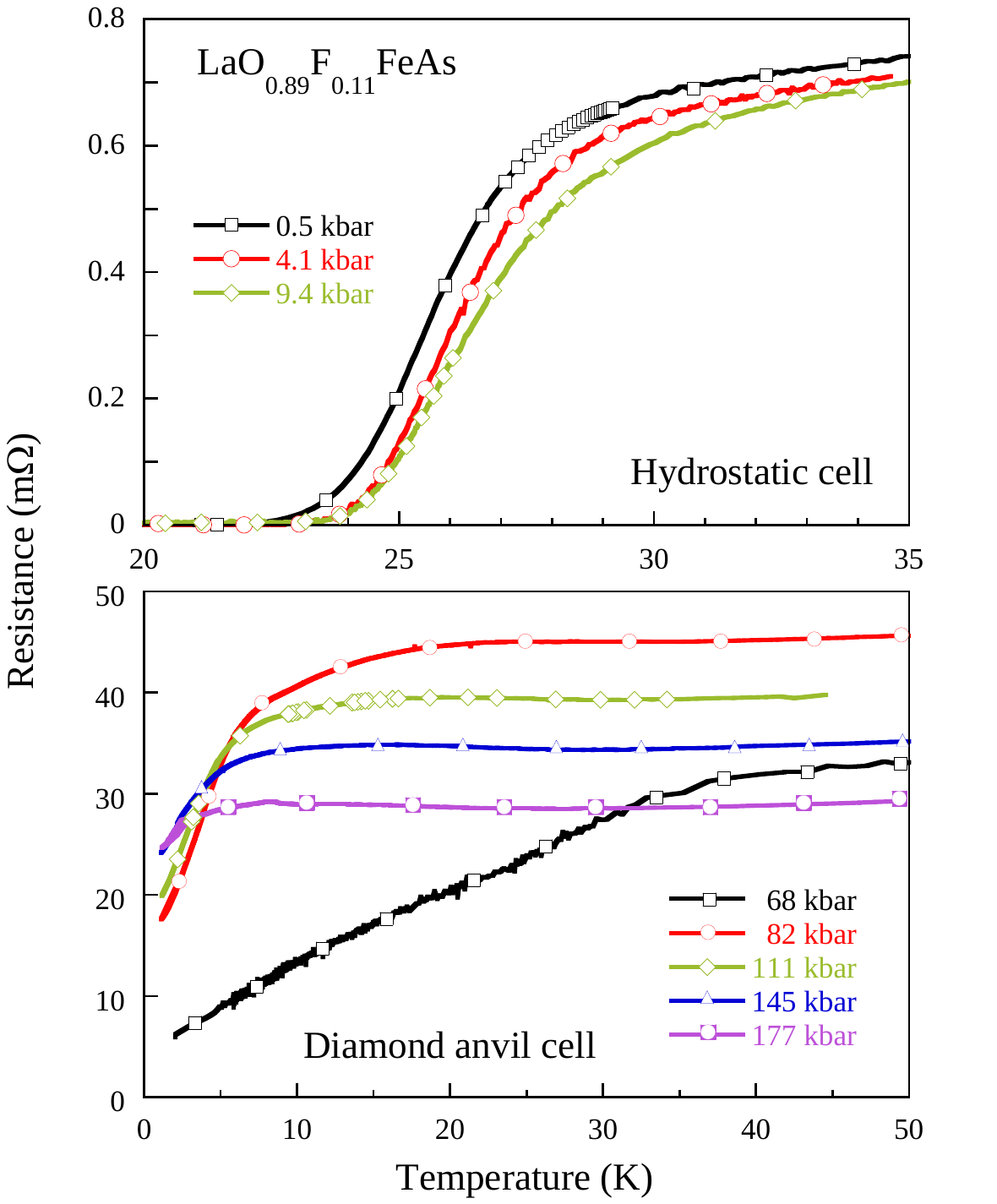}
    \end{center}
    \caption{(color) Electrical resistance of La[O$_{0.89}$F$_{0.11}$]FeAs for different pressures.
    Upper panel: curves corresponding to the hydrostatic clamp. Lower panel: diamond anvil cell experiment.}
    \label{fig:fig01}
\end{figure}

\section{La[O$_{0.89}$F$_{0.11}$]FeAs}
We performed two separate high pressure-experiments on
polycrystalline samples of La[O$_{0.89}$F$_{0.11}$]FeAs. The samples
were taken from the same batch described in Ref.
\cite{sales_2008_1}. The first experiment was performed in a Teflon
capsule piston-cylinder cell and utilized a nearly hydrostatic 1:1
mixture of $n$-pentane:isoamyl alcohol as the pressure medium. Pressure
was increased at room temperature and determined at low temperature
using the superconducting transition of a piece of Pb located next
to the sample \cite{wittig_1988_1}. Electrical resistivity data were
obtained using a 4-lead technique with a LR-700 AC resistance bridge
at pressures of 0.5, 4.1 and 9.4 kbar. Figure \ref{fig:fig01} (top)
shows the temperature dependence of the resistance for each pressure
measured in the hydrostatic cell. Upon attempting to increase the
pressure beyond 9.4 kbar, the cell failed and we continued the
experiment in a diamond anvil cell (DAC).

\begin{figure}[h]
    \begin{center}
        \includegraphics[width=8cm]{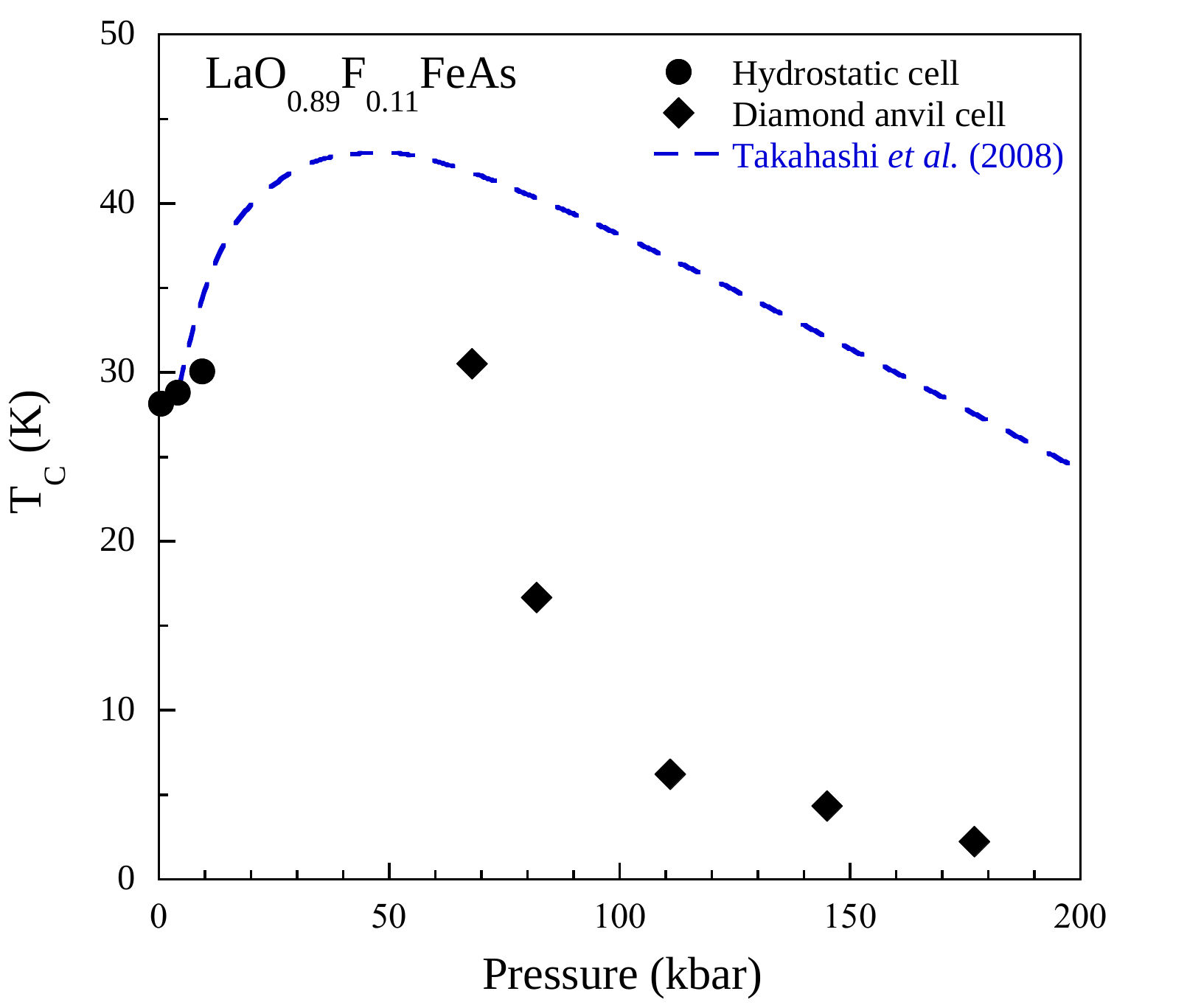}
    \end{center}
    \caption{(color) Superconducting $T_c$ vs. pressure for
    La[O$_{0.89}$F$_{0.11}$]FeAs. Circles correspond to measurements in the hydrostatic cell,
    while diamonds correspond to the diamond anvil cell experiment.
    The dashed curve represents the data previously published by Takahashi \textit{et al.} \cite{takahashi_2008_1}.}
    \label{fig:fig02}
\end{figure}

The DAC is a mechanically loaded commercial model, manufactured by Kyowa Seisakusho Ltd. The diamond anvils are beveled from 500 to 250 $\mu$m tips.  One of the diamonds contains six deposited tungsten microprobes encapsulated in high-quality homepitaxial diamond. The fabrication of ``designer'' diamonds is described in Ref. \cite{weir_2000_1}. The gasket was made from a 200 $\mu$m thick MP35N foil preindented to 40-50 $\mu$m and a 100 $\mu$m diameter hole was drilled through the gasket using an electrical discharge machine (EDM). Several $\sim$ 5 $\mu$m diameter ruby spheres were loaded into the hole in the gasket and the remaining space in the hole was filled with several small chunks of polycrystalline La[O$_{0.89}$F$_{0.11}$]FeAs. Pressure was adjusted and determined at room temperature, using the fluorescence spectrum of the ruby spheres and the calibration of Chijioke \textit{et al.} \cite{chijioke_2005_1}. Further details of the DAC technique are described in Ref. \cite{jeffries_2006_1}.

Figure \ref{fig:fig01} (bottom) shows the temperature dependence of the resistivity as measured in the DAC.  Following the initial measurement at 68 kbar, pressure was increased to 177 kbar and then decreased to 145, 111 and 82 kbar.  Note that the sample is in direct contact with the metallic gasket so that the measured resistance results from a combination of the sample and gasket resistivity.  However, such a configuration is sufficient for locating the sharp drop in resistance when the sample becomes superconducting.  The critical temperature was determined by the temperature at which the sample resistance reached 90\% of the normal state value just above $T_c$.

Figure \ref{fig:fig02} illustrates the pressure dependence of $T_c$
for La[O$_{0.89}$F$_{0.11}$]FeAs.  For comparison, the
results of Takahashi \textit{et al.} \cite{takahashi_2008_1} are
plotted as a dashed line. While $T_c$ initially increases and then
decreases beyond $\sim$ 68 kbar, bearing some resemblance to the
previous measurements, the magnitude of the superconducting
transition temperatures are quite different. The differing pressure
dependencies could be due to differences in stoichiometry or sample
purity. The reported fluorine concentrations are nominal
concentrations determined by the stoichiometry of the unreacted
mixtures. The final concentration of fluorine in the reacted
material may depend strongly on the details of the high-temperature
processing such as the annealing schedule.

\section{Ce[O$_{0.88}$F$_{0.12}$]FeAs}
We performed two high pressure experiments on polycrystalline samples of Ce[O$_{0.88}$F$_{0.12}$]FeAs prepared at Oak Ridge National Laboratory.  The compound was synthesized from CeAs, Fe$_2$O$_3$, Fe, As, and a 1:1 mixture of Ce and CeF$_3$, combined to produce a nominal composition of Ce[O$_{0.88}$F$_{0.12}$]FeAs. The reactants (3 g total mass) were intimately mixed and pressed into a pellet, sealed in an silica tube under about $1/3$ atm of Ar, and heated for 36 hours at 1200 C. Reaction with vapor in the tube produces a thin light gray coating on the surface of the pellet which was gently sanded away. The sample was then re-pelletized and heated for an additional 12 hours under the same conditions. The thin surface-layer was sanded away prior to characterization. Powder X-ray diffraction analysis revealed two small impurity phases, FeAs and CeAs, present at the level of a few percent.

\begin{figure}[h]
    \begin{center}
        \includegraphics[width=8cm]{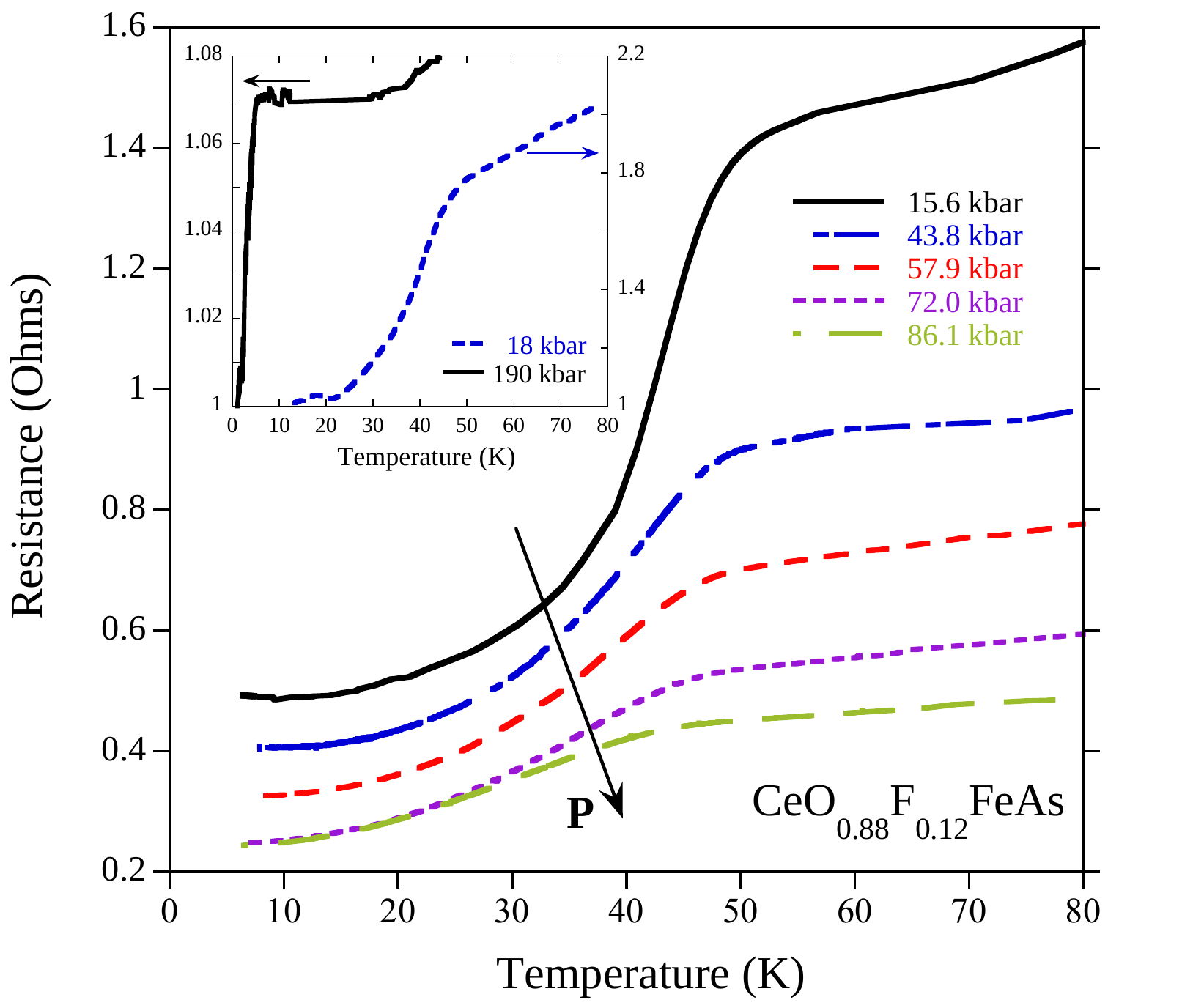}
    \end{center}
    \caption{(color) Temperature dependence of the electrical resistance of
Ce[O$_{0.88}$F$_{0.12}$]FeAs for different applied pressures up to
86 kbar using the Bridgman anvil cell technique. Here we have also
defined the superconducting $T_c$ as the temperature at which the
sample resistance reaches 90\% of the normal state value.
\textit{Inset}: diamond anvil cell resistance curves for 18 and 190
kbar, normalized to their respective base temperature values.}
    \label{fig:fig03}
\end{figure}

At ambient pressure, the sample displays $T_c$ values of 40 and 44 K in magnetic susceptibility and resistivity, respectively. The resistive $T_c$ value is determined by the temperature at which the resistivity drops to 90\% of the normal state value just above $T_c$. The first high-pressure experiment was performed in a Bridgman anvil cell with 4 mm diameter anvil flats and utilized quasi-hydrostatic solid steatite as pressure medium. Pressure is applied at room temperature using a hydraulic press and clamped with a nut. At pressures up to 44 kbar, the pressure was determined using a Sn manometer located next to the sample and the calibration of Smith \textit{et al.} \cite{smith_1969_1}. Above 44 kbar, the wires connecting to the Sn manometer broke and pressure was determined from the applied load. Our previous experience with these Bridgman anvil cells indicates that the pressure determined from the applied load is accurate to within $\pm$ 20\%.  Figure \ref{fig:fig03} shows the results of the Bridgman cell measurements on Ce[O$_{0.88}$F$_{0.12}$]FeAs.

\begin{figure}[h]
    \begin{center}
        \includegraphics[width=8cm]{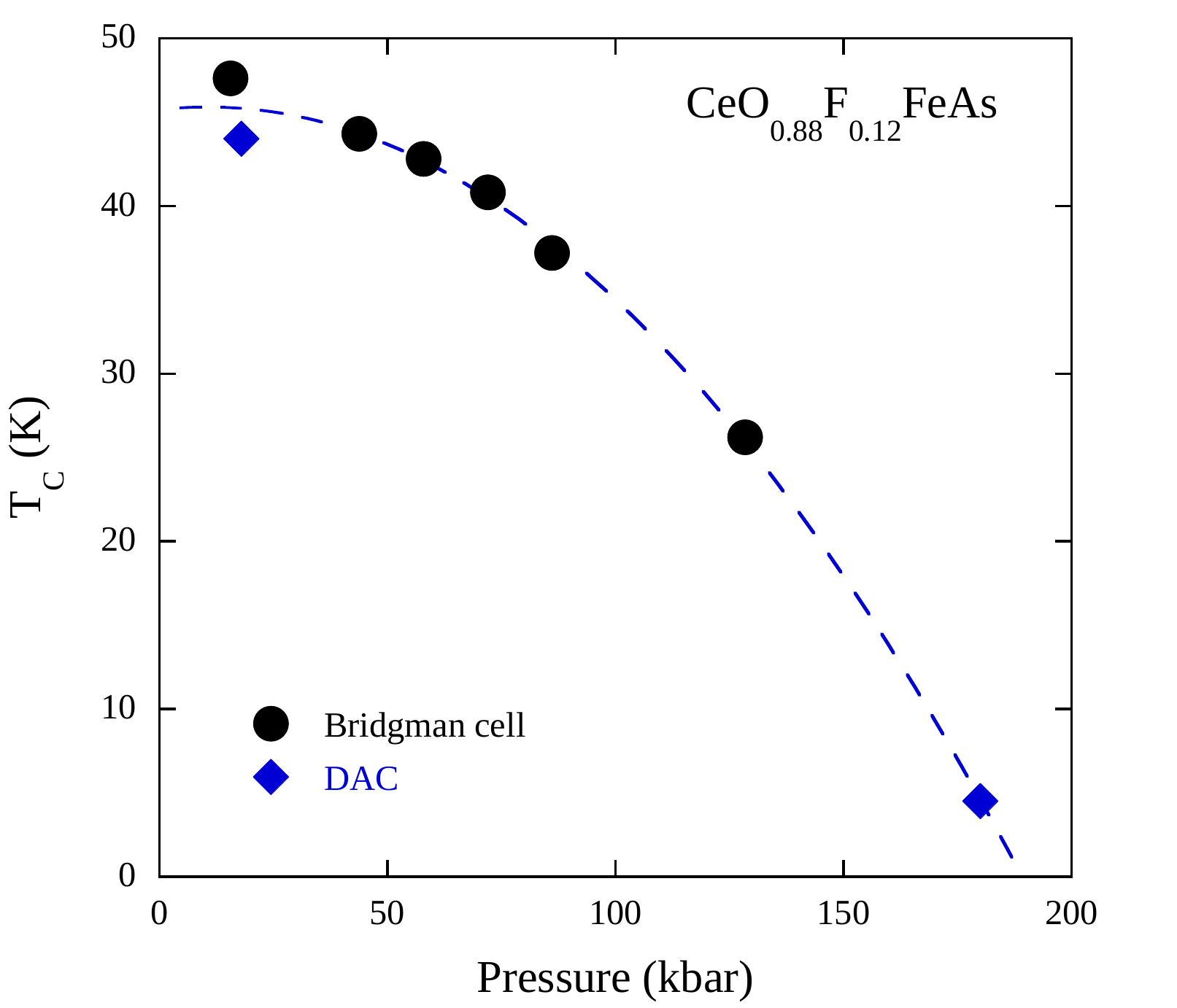}
    \end{center}
    \caption{(color) Superconducting $T_c$ versus pressure for Ce[O$_{0.88}$F$_{0.12}$]FeAs. The dashed curve is a guide to the eye given by a quadratic fit of the data, extrapolating to zero temperature at $\sim$ 190 kbar.}
    \label{fig:fig04}
\end{figure}

In order to verify the pressure dependence of $T_c$ and to extend
the results to higher pressure, we performed a second experiment on
Ce[O$_{0.88}$F$_{0.12}$]FeAs using a DAC. The technique used was the
same as that described for the lanthanum-based compound. Figure \ref{fig:fig03}
(\textit{inset}) shows resistance versus temperature obtained
from the DAC measurements. Figure \ref{fig:fig04} shows the $T_c$
versus pressure for Ce[O$_{0.88}$F$_{0.12}$]FeAs
obtained from the Bridgman and DAC experiments.  At the lowest measured pressures we find $T_c$ values near 47 K, somewhat higher than the $T_c=44$ K measured at ambient pressure.  This small increase in the critical temperature is likely due to improvement in sample connectivity due to grain compaction under pressure, rather than the intrinsic pressure dependence of $T_c$.  The $T_c$ value decreases monotonically with pressure and is suppressed below 1.1 K at 265 kbar.

The strong dependence of $T_c$ on pressure in these materials is
rather remarkable.  Shein \textit{et al.} \cite{shein_2008_1}
calculate that the bulk modulus of LaOFeAs is only 98 GPa,
significantly smaller than that found for the cuprate
superconductors. It is likely that the strong dependence of $T_c$ on
pressure for La[O$_{0.89}$F$_{0.11}$]FeAs and
Ce[O$_{0.88}$F$_{0.12}$]FeAs is related to their high
compressibility. Experiments to determine structural parameters
under pressure would help to clarify the effect of lattice
properties on $T_c$.

For the oxypnictides, it is likely that, as in the cuprates, increasing
pressure leads to an increase in carrier concentration.  The initial
increase in $T_c$ with pressure for La[O$_{0.89}$F$_{0.11}$]FeAs may
thus be due to the sample being underdoped.  Indeed, Lu
\textit{et al.} \cite{lu_2008_1} find that increased doping achieved
through high-pressure synthesis raises $T_c$ to 41 K in
La[O$_{0.4}$F$_{0.6}$]FeAs.  In the high-$T_c$ cuprate
superconductors, it is found that $T_c$ generally increases with
pressure in optimally doped samples, highlighting the fact that the
effect of pressure is more complicated than simply changing the
carrier concentration \cite{schilling_2001_1}.  The negative
pressure dependence of $T_c$ that we find for apparently optimally
doped Ce[O$_{1-x}$F$_x$]FeAs combined with that previously reported
for Sm[O$_{1-x}$F$_x$]FeAs \cite{lorenz_2008_1} points to a
possible difference between the oxypnictide and cuprate
superconductors.  A systematic study of the effect of pressure on
$T_c$ across a wide range of dopings is clearly needed in order to
obtain a better understanding of the optimal conditions for
high-$T_c$ values in the oxypnictide superconductors.

\textbf{Acknowledgements}

Research at University of California, San Diego, was supported by the National Nuclear Security Administration under the Stewardship Science Academic Alliance Program through the U.S. Department of Energy grant number DE-FG52-06NA26205.  Research at Oak Ridge National Laboratory is sponsored by the Division of Materials Sciences and Engineering, Office of Basic Energy Sciences.  Oak Ridge National Laboratory is managed by UT-Battelle, LLC, for the U.S. Department of Energy.  Lawrence Livermore National Laboratory is operated by Lawrence Livermore National Security, LLC, for the U.S. Department of Energy, National Nuclear Security Administration under Contract DE-AC52-07NA27344.  YKV acknowledges support from the Department of Energy (DOE) – National Nuclear Security Administration (NNSA) under Grant No. DE-FG52-06NA26168.




\bibliographystyle{elsart-num}
\bibliography{oxypnictide_pressure_3}

\end{document}